\documentclass[apj]{emulateapj}
\usepackage{apjfonts}

\usepackage[usenames,dvips]{color}

\begin{document}
\journalinfo{ApJ, accepted}
%

\shorttitle{Polarimetric VLBI Methods for Sgr A* Flares}

\shortauthors{Fish et al.}

\title{Detecting Changing Polarization Structures in
  Sagittarius~A* with High Frequency VLBI}
\author{Vincent L.~Fish\altaffilmark{1},
        Sheperd S.~Doeleman\altaffilmark{1},
        Avery E.~Broderick\altaffilmark{2},
        Abraham Loeb\altaffilmark{3}, \&
        Alan E.~E.~Rogers\altaffilmark{1}}

\altaffiltext{1}{Massachusetts Institute of Technology, Haystack
  Observatory, Route 40, Westford, MA 01886.}

\altaffiltext{2}{Canadian Institute for Theoretical Astrophysics,
  University of Toronto, 60 St.\ George St., Toronto, ON M5S 3H8,
  Canada.}

\altaffiltext{3}{Institute for Theory and Computation, Harvard
  University, Center for Astrophysics, 60 Garden St., Cambridge, MA
  02138.}
\begin{abstract}
Sagittarius~A* is the source of near infrared, X-ray, radio, and
(sub)millimeter emission associated with the supermassive black hole
at the Galactic Center.  In the submillimeter regime, Sgr~A* exhibits
time-variable linear polarization on timescales corresponding to $<
10$ Schwarzschild radii of the presumed $4 \times 10^6$~M$_\sun$ black
hole.  In previous work, we demonstrated the potential for
total-intensity (sub)millimeter-wavelength very long baseline
interferometry (VLBI) to detect time-variable -- and periodic --
source structure changes in the Sgr~A* black hole system using
nonimaging analyses.  Here we extend this work to include full
polarimetric VLBI observations.  We simulate full-polarization
(sub)millimeter VLBI data of Sgr~A* using a hot-spot model that is
embedded within an accretion disk, with emphasis on nonimaging
polarimetric data products that are robust against calibration errors.
Although the source-integrated linear polarization fraction in the
models is typically only a few percent, the linear polarization
fraction on small angular scales can be much higher, enabling the
detection of changes in the polarimetric structure of Sgr~A* on a wide
variety of baselines.  The shortest baselines track the
source-integrated linear polarization fraction, while longer baselines
are sensitive to polarization substructures that are beam-diluted by
connected-element interferometry.  The detection of periodic
variability in source polarization should not be significantly
affected even if instrumental polarization terms cannot be calibrated
out.  As more antennas are included in the (sub)mm-VLBI array,
observations with full polarization will provide important new
diagnostics to help disentangle intrinsic source polarization from
Faraday rotation effects in the accretion and outflow region close to
the black hole event horizon.
\end{abstract}

\section{Introduction}

The Galactic Center source Sagittarius~A* (Sgr~A*) provides the best
case for high-resolution, detailed observations of the accretion and
outflow region surrounding the event horizon of a black hole.  There
are several compelling reasons to observe Sgr~A* with very long
baseline interferometry (VLBI) at (sub)millimeter\footnote{We shall
  henceforth use the term ``millimeter'' to denote wavelengths of
  1.3~mm of shorter (in contrast with observations at 3~mm and 7~mm,
  which are sometimes also referred to as ``millimeter''
  wavelengths).} wavelengths.  The spectrum of Sgr~A* peaks in the
millimeter \citep[][and references therein]{markoff07}.  Interstellar
scattering, which varies as the wavelength $\lambda^2$, becomes less
than the fringe spacing of the longest baseline available to VLBI in
the millimeter-wavelength regime.  Indeed, VLBI on the longest
baselines available at 345~GHz probes scales of twice the
Schwarzschild radius ($R_\mathrm{S}$) for a $4 \times 10^6$~M$_\sun$
black hole.  From previous observations at 230~GHz, it is known that
there are structures on scales smaller than a few $R_\mathrm{S}$
\citep{doeleman08}.  Such high angular resolution, presently
unattainable by any other method (including facility instruments such
as the Very Long Baseline Array), is necessary to match the expected
spatial scales of the emitting plasma in the innermost regions
surrounding the black hole and will be required to unambiguously
determine the inflow/outflow morphology and permit tests of general
relativity.

This sensitivity to small spatial scales also makes millimeter
polarimetric VLBI possible.  Although the linear polarization fraction
of Sgr~A* integrated over the entire source is only a few percent
\citep[e.g.,][]{marrone07}, the fractional polarization on small
angular scales is likely much larger.  In general, relativistic
accretion flow models predict that the electric vector polarization
angle (EVPA) will vary along the circumference of the accretion disk
\citep{bromley01,broderick05,broderick06}, indicating that single-dish
observations and connected-element interferometers probably
underestimate linear polarization fractions due to beam
depolarization.

Polarized synchrotron radiation coming from Sgr~A* was detected by
\citet{aitken00} at millimeter and submillimeter wavelengths.
Multiple observations since then have demonstrated that the polarized
emission is variable on timescales from hours to many days
\citep{bower05,macquart06,marrone06a,marrone07,marrone08}.  In one
case, the timescale of variability and the trace of polarization in
the Stokes $(Q,U)$ plane of the millimeter-wavelength emission are
suggestive of the detection of an orbit of a polarized blob of
material \citep{marrone06b}.  Near infrared observations by
\citet{trippe07} are also consistent with a hot spot origin for
periodic variability.  It is possible that connected-element
interferometry may suffice to demonstrate polarization periodicity,
but millimeter-wavelength VLBI, which effectively acts as a spatial
filter on scales of a few to a few hundred $R_\mathrm{S}$, can be more
sensitive to changing polarization structures.

Initial millimeter VLBI observations of Sgr~A* will necessarily
utilize non-imaging analysis techniques, for reasons outlined in
\citet[][henceforth Paper~I]{doeleman09}.  One way to do this is to
analyze so-called interferometric ``closure quantities,'' which are
relatively immune to calibration errors \citep{rogers74,rogers95}.  In
Paper~I, we considered prospects for detecting the periodicity
signature of a hot spot orbiting the black hole in Sgr~A* via closure
quantities in total-intensity millimeter-wavelength VLBI.  In the
single polarization case, it is necessary to construct closure
quantities from at least three or four antennas in order to produce
robust observables, since the timescales of atmospheric coherence and
frequency standard stability do not permit standard nodding
calibration techniques.  Closure quantities can be used in
full-polarization observations as well, but it is also possible to
construct robust observables on a baseline of two antennas by taking
visibility ratios between different correlation products.  In this
work, we extend our techniques to explore polarimetric signatures of a
variable source structure in Sgr~A*, with emphasis on ratios of
baseline visibilities.

\section{Models and Methods}
\label{models}

We employ the same models discussed in Paper~I to describe the flaring
emission of Sgr~A*, and shall only briefly review these here,
directing the reader to Paper~I, and references therein, for more
detail.  These models consist of an orbiting hot spot, modelled by a
gaussian over-density of power-law electrons, embedded in a
radiatively inefficient accretion flow, containing both thermal and
non-thermal electron populations.

The primary emission mechanism for both components is synchrotron.  We
model the emission from the thermal and nonthermal electrons using the
emissivities described in \citet{yuan03} and \citet{jones77},
respectively, appropriately modified to account for relativistic
effects (see \citealt{broderick04} for a more complete description of
polarized general relativistic radiative transfer).  Since we
necessarily are performing the fully polarized radiative transfer, for
the thermal electrons we employ the polarization fraction derived in
\citet{petrosian83}.  In doing so we have implicitly assumed that the
emission due to the thermal electrons is locally isotropic, which,
while generally not the case in the presence of ordered magnetic
fields, is unlikely to modify our results significantly.  For both
electron populations the absorption coefficients are determined
directly via Kirchoff's law.

As described in Paper I, the assumed magnetic field geometry was
toroidal, consistent with simulations and analytical expectations for
magnetic fields in accretion disks, though other field configurations
are possible \citep[e.g.,][]{huang09}.  While the overall flux of our
models is relatively insensitive to the magnetic field geometry, the
polarization is dependent on it.  However, polarization light curves
and maps with considerably different magnetic field geometries (e.g.,
poloidal) are qualitatively similar, showing large swings in
polarization angle and patches of nearly uniform polarization in the
images.

Generally, synchrotron emission has both linearly and circularly
polarized components.  However, the circular polarization fraction is
suppressed by an additional factor of the electron Lorentz factor.
For the electrons producing the millimeter emission, this corresponds
to a reduction by a factor of $30$--$100$ in Stokes $V$ in comparison
to Stokes $Q$ and $U$.  This is consistent with observations by
\citet{marrone06a}, who obtain an upper limit of $\sim 1\%$ circular
polarization at 340~GHz.  Therefore, we explicitly omitted the
circular polarization terms in the computation of flaring
polarization.

In addition, we have neglected the potentially modest intrinsic
Faraday rotation.  Within $r\lesssim 10^2$--$10^3R_S$ the accreting
electrons are expected to be substantially relativistic, and thus not
contribute significantly to the rotation measure within the
millimeter-emitting region.  This is consistent with the lack of
observed Faraday depolarization at these wavelengths
\citep[e.g.,][]{aitken00,marrone07}, which itself implies the absence
of significant in situ Faraday rotation.  Similarly, beam
depolarization caused by variations within an external Faraday screen
on angular scales comparable to that of the emission region are
empirically excluded.  This leaves the possibility of a smoothly
varying external Faraday screen, which manifests itself in the VLBI
data as an additional phase difference between right and left
circularly polarized visibilities, but does not affect our analysis
otherwise.

\begin{figure}
\resizebox{\hsize}{!}{\includegraphics{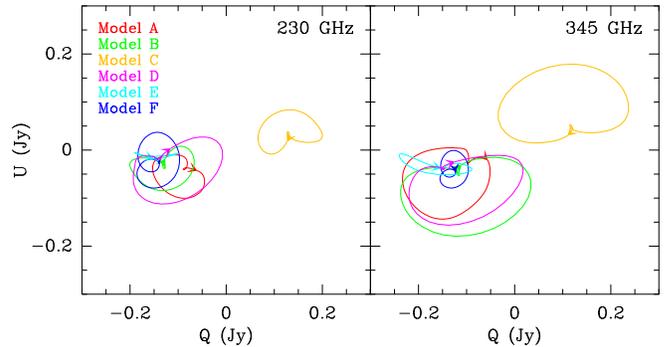}}
\caption{Integrated polarization traces of the models in the Stokes
  $(Q,U)$ plane at 230 and 345~GHz over a full hot spot orbit, as
  would be seen by the SMA (for instance).
\label{fig-qu}
}
\end{figure}

Model images are created in each of the Stokes parameters $I$, $Q$,
and $U$.  Six models differing in hot spot orbital period, black hole
spin, and accretion disk inclination and major-axis orientation are
produced at 230 and 345~GHz, as in Paper~I.  Model properties are
summarized in Table~\ref{tab-models}.  Source-integrated linear
polarization fractions range from 0.8 to 26\% for models including
both a disk and a hot spot, depending on the model and hot spot
orbital phase, with typical integrated quiescent polarization
fractions (of the disk alone) of 10 to 15\%.  Integrated EVPA
variation over the course of the hot spot orbit ranges from 4 to
57\degr, depending on the model.  The integrated polarization
fractions and EVPA variations as well as the polarization traces in
the Stokes $(Q,U)$ plane (Fig.~\ref{fig-qu}) are all broadly
consistent with the range of variability seen in the Submillimeter
Array (SMA) observations reported by \citet{marrone06b}.  The local
linear polarization fraction can be much higher, exceeding 70\% in
some parts of the accretion disk.

Simulated array data are produced by the Astronomical Image Processing
System (AIPS) task UVCON for each of the Stokes parameters.  The array
is taken to consist of up to seven stations: the Caltech Submillimeter
Observatory, James Clerk Maxwell Telescope, and six SMA telescopes
phased together into a single station (Hawaii); the Arizona Radio
Observatory Submillimeter Telescope (SMT); a phased array consisting
of eight telescopes in the Combined Array for Research in
Millimeter-wave Astronomy (CARMA); the Large Millimeter Telescope
(LMT); the 30 m Institut de radioastronomie millim\'{e}trique dish at
Pico Veleta (PV); the Plateau de Bure Interferometer phased together
as a single station (PdB); and a site in Chile, either a single 10 or
12 m class telescope (Chile 1) or a phased array of 10 dishes of the
Atacama Large Millimeter Array (Chile 10).  Details of the method as
well as assumed parameters of the telescopes are given in Paper I.

\begin{deluxetable*}{llrcrcccccrcc}
\tabletypesize{\footnotesize}
\tablecaption{Model Parameters\label{tab-models}}
\tablehead{
  \colhead{} &
  \colhead{$a$} &
  \colhead{Period} &
  \colhead{$i$} &
  \colhead{PA\tablenotemark{b}} &
  \colhead{$\nu$} &
  \colhead{Disk Flux\tablenotemark{c}} &
  \colhead{Min Flux\tablenotemark{c}} &
  \colhead{Max Flux\tablenotemark{c}} &
  \colhead{Disk Pol.\tablenotemark{d}} &
  \colhead{Disk EVPA\tablenotemark{d}} &
  \colhead{Min Pol.\tablenotemark{d}} &
  \colhead{Max Pol.\tablenotemark{d}}
\\
  \colhead{Model} &
  \colhead{($R_G$)\tablenotemark{a}} &
  \colhead{(min)} &
  \colhead{(\degr)} &
  \colhead{(\degr)} &
  \colhead{(GHz)} &
  \colhead{(Jy)} &
  \colhead{(Jy)} &
  \colhead{(Jy)} &
  \colhead{(\%)} &
  \colhead{(\degr)} &
  \colhead{(\%)} &
  \colhead{(\%)} 
}
\startdata
 A & 0   & 27.0 & 30 & 90 & 230 & 3.19 & 3.49 & 4.05 & 10 & $-$75 & 7.0 & 16 \\
   &     &      &    &    & 345 & 3.36 & 3.63 & 5.28 & 11 & $-$81 & 6.0 & 24 \\
 B & 0   & 27.0 & 60 & 90 & 230 & 3.03 & 3.05 & 4.03 & 14 & $-$84 & 6.9 & 20 \\
   &     &      &    &    & 345 & 2.96 & 2.99 & 4.78 & 13 & $-$80 & 2.2 & 26 \\
 C & 0   & 27.0 & 60 &  0 & 230 & 3.03 & 3.05 & 4.03 & 14 &     6 & 6.9 & 20 \\
   &     &      &    &    & 345 & 2.96 & 2.99 & 4.78 & 13 &    10 & 2.2 & 26 \\
 D & 0.9 & 27.0 & 60 & 90 & 230 & 2.98 & 2.99 & 4.05 & 15 & $-$86 & 0.8 & 21 \\
   &     &      &    &    & 345 & 2.96 & 2.97 & 4.00 & 15 & $-$82 & 1.6 & 24 \\
 E & 0.9 &  8.1 & 60 & 90 & 230 & 2.98 & 3.08 & 4.15 & 15 & $-$86 & 10  & 19 \\
   &     &      &    &    & 345 & 2.96 & 3.04 & 6.07 & 15 & $-$82 & 9.7 & 24 \\
 F & 0   &166.9 & 60 & 90 & 230 & 3.07 & 3.08 & 3.38 & 15 & $-$84 & 9.8 & 19 \\
   &     &      &    &    & 345 & 2.99 & 3.00 & 3.18 & 13 & $-$80 & 10  & 17
\enddata
\tablenotetext{a}{Spin is given in units of the gravitational radius,
  $R_G \equiv GMc^{-2} = \onehalf\,R_\mathrm{S}$.}
\tablenotetext{b}{Accretion disk major axis position angle (east of
  north).}
\tablenotetext{c}{Stokes $I$ flux density of integrated quiescent disk
  alone and minimum/maximum of system with orbiting hot spot.}
\tablenotetext{d}{Polarization fraction and EVPA of the disk emission
  alone and minimum/maximum polarization fraction of system with
  orbiting hot spot.}
\end{deluxetable*}

\pagebreak

\section{Polarimetric Considerations}
\label{polarimetric}

For ideal circularly-polarized feeds, the perfectly calibrated
correlations are related to the complex Stokes visibilities
($I_\nu,Q_\nu,U_\nu,V_\nu$) as follows:
\begin{eqnarray*}
RR & = & I_\nu + V_\nu \\
LL & = & I_\nu - V_\nu \\
RL & = & Q_\nu + iU_\nu \\
LR & = & Q_\nu - iU_\nu,
\end{eqnarray*}
where $i = \sqrt{-1}$ and $RL$ (for example) denotes the right
circular polarized signal at one station correlated against the left
circular polarized signal at another.  We have used the convention of
\citet{cotton93}.  Other definitions, differing in sign or rotation of
the $RL$ and $LR$ terms by factors of $i$, are possible
\citep[e.g.,][]{thompson01}, but do not affect the analysis.
Significant circular polarization is neither predicted in the hot spot
models nor observed at the resolution of connected-element arrays
\citep{bower03,marrone06a}.  In the limit of no circular polarization
($V_\nu = 0$), $I_\nu = RR = LL$ is a direct observable in the
parallel-hand correlations, but $Q_\nu$ and $U_\nu$ appear only in
combination in the cross-hand correlations.  $RL$ and $LR$
visibilities, which are direct observables, are constructed by
appropriate complex addition of the Stokes $Q_\nu$ and $U_\nu$
visibilities.  Right- and left-circular polarized (RCP and LCP) feeds
are preferable to linearly-polarized feeds for detecting linear
polarization, since the latter mix Stokes $I_\nu$ with $Q_\nu$ in the
parallel-hand correlations \citep{thompson01}.

For a point source, $I_\nu \geq \sqrt{Q_\nu^2 + U_\nu^2 + V_\nu^2}$.
However, for an extended distribution, the polarized Stokes
visibilities can exceed the amplitude of the Stokes $I_\nu$
visibility.  (For instance, a uniform total intensity distribution
with constant linear polarization fraction but a changing linear
polarization angle will produce no power in Stokes $I_\nu$ on scales
small compared to the distribution, but the Stokes visibilities
$Q_\nu$ and $U_\nu$ will be nonzero.)

Analysis of polarimetric data is more complex than total intensity
(Stokes $I$; we will henceforth drop the subscript on Stokes
visibilities) data, but ratios of cross-hand ($RL$ and $LR$) to
parallel-hand ($RR$ and $LL$) visibilities provide robust
baseline-based observables immune to most errors arising from
miscalibrated antenna complex gains.  This stands in contrast to the
single-polarization case in which robust observables can only be
constructed from closure quantities on three or more telescopes.  The
procedure for referencing cross-hand data to parallel-hand data is
explained in detail in \citet{cotton93} and \citet{roberts94} and has
been used successfully in experiments \citep[e.g.,][]{wardle71}.
Several details warrant further discussion.  We shall refer to the
full expressions for the observed correlation quantities:
\begin{eqnarray*}
RR = R_1R_2^*  =  G_{1R}G_{2R}^* & [ &
                        (I_{12}+V_{12}) e^{i(-\varphi_1+\varphi_2)} \\
  & & + D_{1R} D_{2R}^* (I_{12}-V_{12}) e^{i(+\varphi_1-\varphi_2)} \\
  & & + D_{1R}           P_{21}^*       e^{i(+\varphi_1+\varphi_2)} \\
  & & + D_{2R}^*         P_{12}         e^{i(-\varphi_1-\varphi_2)}], \\
LL = L_1L_2^*  =  G_{1L}G_{2L}^*  & [ &
                        (I_{12}-V_{12}) e^{i(+\varphi_1-\varphi_2)} \\
  & & + D_{1L} D_{2L}^* (I_{12}+V_{12}) e^{i(-\varphi_1+\varphi_2)} \\
  & & + D_{1L}           P_{12}         e^{i(-\varphi_1-\varphi_2)} \\
  & & + D_{2L}^*         P_{21}^*       e^{i(+\varphi_1+\varphi_2)}], \\
RL = R_1L_2^*  =  G_{1R}G_{2L}^*  & [ &
                         P_{12}         e^{i(-\varphi_1-\varphi_2)} \\
  & & + D_{1R} D_{2L}^*  P_{21}^*       e^{i(+\varphi_1+\varphi_2)} \\
  & & + D_{1R}          (I_{12}-V_{12}) e^{i(+\varphi_1-\varphi_2)} \\
  & & + D_{2L}^*        (I_{12}+V_{12}) e^{i(-\varphi_1+\varphi_2)}], \\
LR = L_1R_2^*  =  G_{1L}G_{2R}^*  & [ &
                         P_{21}^*       e^{i(+\varphi_1+\varphi_2)} \\
  & & + D_{1L} D_{2R}^*  P_{12}         e^{i(-\varphi_1-\varphi_2)} \\
  & & + D_{1L}          (I_{12}+V_{12}) e^{i(-\varphi_1+\varphi_2)} \\
  & & + D_{2R}^*        (I_{12}-V_{12}) e^{i(+\varphi_1-\varphi_2)}],
\end{eqnarray*}
where numeric subscripts refer to antenna number, letter subscripts
refer to the polarization (RCP or LCP), a star denotes complex
conjugation, $G_{nX} = g_{nX} e^{i\psi_{nX}}$ is the complex gain in
polarization $X \in \{R,L\}$ at antenna $n$, $P = Q + iU$, $D_{nX}$ is
the instrumental polarization, and $\varphi_n$ is the parallactic
angle \citep[equations reproduced from][]{roberts94}.  The $\varphi_n$
terms are constant for equatorial mount telescopes and can be
incorporated into the $G$ and $D$ terms, while for alt-azimuth mount
telescopes the $\varphi_n$ terms vary predictably based on source
declination, hour angle, and antenna latitude.  It is likely that all
of the telescopes in potential millimeter-wavelength VLBI arrays in
the near future will have $\varphi_n$ terms varying with parallactic
angle.

\begin{figure}[t]
\resizebox{\hsize}{!}{\includegraphics{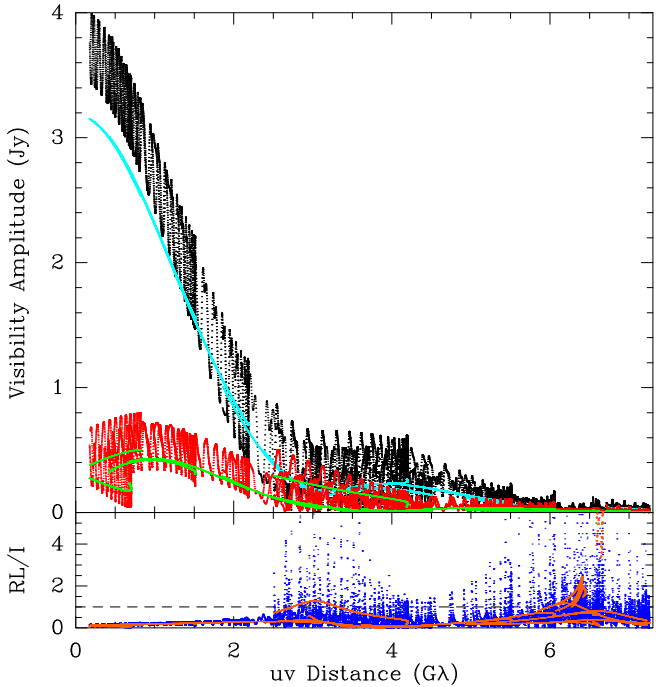}}
\caption{\emph{Top}: Visibility amplitude as a function of projected
  baseline length ($\sqrt{u^2+v^2}$) for Model~A at 230~GHz
  (noiseless).  Stokes $I$ is shown in black, and $RL$ is shown in
  red.  A real orbiting hot spot would persist for only a small
  fraction of a day, producing a plot corresponding to a subset of the
  above points.  Contributions from the disk alone in the absence of a
  hot spot are shown in cyan (Stokes $I$) and green
  ($RL$). \emph{Bottom}: Ratio of $RL/I$ visibility amplitudes for the
  disk and hot spot (blue) and disk alone (orange).  On small scales,
  $RL/I$ can greatly exceed unity.
\label{fig-uvplt}
}
\end{figure}

The ratio of cross-hand to parallel-hand data (e.g., $RL/LL$) contains
an additional phase contribution $\Psi_n = \psi_{nR} - \psi_{nL}$
equal to the phase difference of the complex gains of the right and
left circular polarizations of antenna $n$ \citep{brown89}.  These
phase differences also enter into closure phases of cross-hand
correlations as $\Psi_1 + \Psi_2 + \Psi_3$.  Fortunately, the
right-left phase differences vary slowly with time
\citep[e.g.,][]{roberts94}, since the atmospheric transmission is not
significantly birefringent at millimeter wavelengths and both
polarizations are usually tied to the same local oscillator.  We will
henceforth assume that the $\Psi$ terms can be properly calibrated
(for instance by observations of an unpolarized calibrator source),
although proper calibration may not be strictly necessary for
periodicity detection, since the expected timescale of variation of
source structure is significantly faster than the timescale of
variation of $\Psi$.  Similarly, it is possible to determine the ratio
of amplitudes of the real gains ($r_n = g_{nR}/g_{nL}$) from
observations of a suitable calibrator.  In general, $r_n$ usually
shows greater short-timescale variability than $\Psi_n$
\citep{roberts94}.  Provided that proper instrumental polarization
calibration is done, the fluctuation in $r_n$ can be estimated from
the $RR/LL$ visibility ratio, since Sgr~A* is expected to have no
appreciable circular polarization\footnote{Stokes $V$ enters the
  expressions for $RR$ and $LL$ only as $I \pm V$, so even if circular
  polarization is detected, it will not prevent estimation of $r_n$
  unless the circular polarization fraction on angular scales
  accessible to VLBI is large or highly variable.}.  Even absent any
complex gain calibration, it is probable that the contamination of the
time series of cross-to-parallel amplitude ratios and (especially)
phase differences by changes in $r_n$ and $\Psi_n$ respectively will
also be seen in the $RR/LL$ amplitude ratio and $RR-LL$ phase
difference.  Thus, large deviations seen in the cross-to-parallel
quantities but not in the parallel-to-parallel quantities will likely
be due to source structure differences, not gain miscalibration.

Correcting for instrumental polarization (the $D$-terms) may be more
difficult.  Effectively, the $D$-terms mix Stokes $I$ into the $RL$
and $LR$ terms \citep{thompson01}.  Observations of calibrators with
the Coordinated Millimeter VLBI Array (CMVA) at $\lambda = 3.5$~mm
found $D$-terms ranging from a few to 21\%, with typical values
slightly greater than 10\% \citep{attridge01,attridge05}.
Polarimetric observations with CARMA and the SMA in their normal
capacity as connected-element interferometers have demonstrated that
the instrumental polarization terms on some of the telescopes that
will be included in future observations may be as low as a few percent
\citep{bower02,marrone06a,marrone07}.  However, it is unknown how
large the $D$-terms will be for potential VLBI arrays at $\lambda =
1.3$ and $0.8$~mm, as many of the critical pieces of hardware
(including feeds, phased-array processors, and even the antennas
themselves) do not yet exist for some of the elements of such arrays.
In any case, contributions from the $D$-terms may be comparable to or
larger than contributions from the source polarization, at least on
the shorter baselines.  The time scale of variations of $D$-terms is
typically much longer than the time scale on which the source
structure in Sgr~A* changes, so carefully-designed observations may
allow for the $D$-terms to be calibrated.  At the angular resolution
of the SMA, polarization fractions of Sgr~A* at 230 and 345~GHz range
between 4 and 10\% \citep{marrone07}, although the polarization
fraction may exceed this range during a flare \citep{marrone08}.
Linear polarization fractions derived from single-dish and
connected-element millimeter observations of Sgr~A* are likely
underestimates of the linear polarization fractions that will be seen
with VLBI, since partial depolarization from spatially separated
orthogonal polarization modes may occur when observed with
insufficient angular resolution to separate them.  That is, the
small-scale structure that will be seen by VLBI is likely to have a
larger polarization fraction than that observed so far with
connected-element interferometery.

Calibration of the electric vector polarization angle (EVPA) may be
difficult, at least in initial observations, due to the lack of known
millimeter-wavelength polarization calibration sources \citep[see,
  e.g.,][]{attridge01}.  EVPA calibration will eventually be important
for understanding the mechanism of linear polarization generation in
Sgr~A*, assuming that the linear polarization can be unambiguously
corrected for Faraday rotation.  However, the ability of cross-hand
correlation data to detect \emph{changes} in the EVPA is unaffected by
absolute EVPA calibration.

In the low signal-to-noise (SNR) regime, the ratio of visibility
amplitudes can be a biased quantity.  Visibility amplitudes are
non-negative by definition, and the complex addition of a large noise
vector to a small signal vector in the visibility plane will bias the
visibility amplitude to higher values.  Nevertheless, even biased
visibility amplitudes may be of some utility in detecting changing
polarization structure.  Since the complex phase of noise is uniform
random, phase differences are unbiased quantities.

\begin{figure*}[p]
\resizebox{\hsize}{!}{\includegraphics{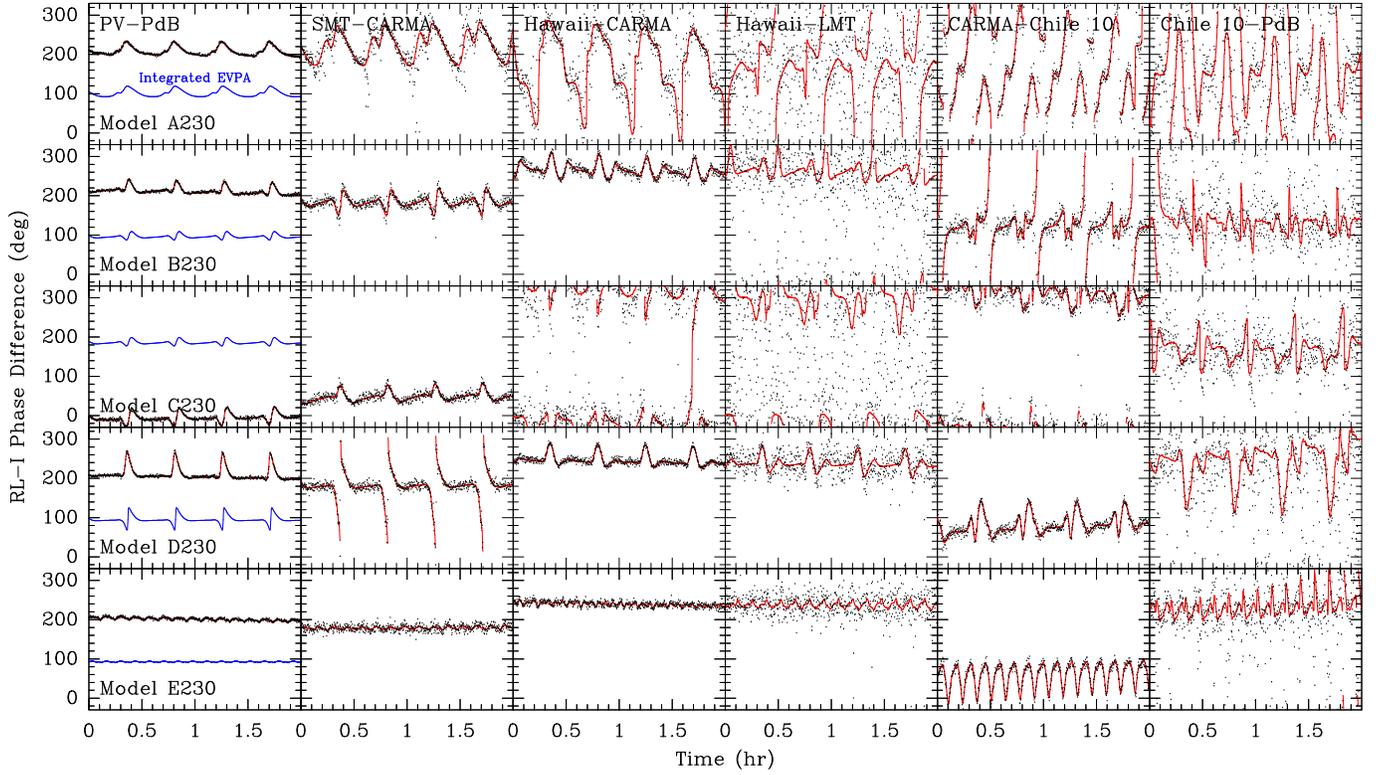}}
\caption{$RL-I$ phase differences for selected baselines at 230~GHz.
  The same 2~hours of data, representing 4.5~periods (14.8~periods for
  Model~E), are shown on all baselines except those involving PV or
  PdB.  The solid line (red in the online edition) indicates the
  expected signal in the absence of noise, and the dots indicate
  simulated data at 8~Gbit\,s$^{-1}$ in each polarization
  (16~Gbit\,s$^{-1}$ total).  The blue line shows the EVPA that would
  be observed if the source were unresolved.
\label{fig-phase230}
}
\end{figure*}

\begin{figure*}[p]
\resizebox{\hsize}{!}{\includegraphics{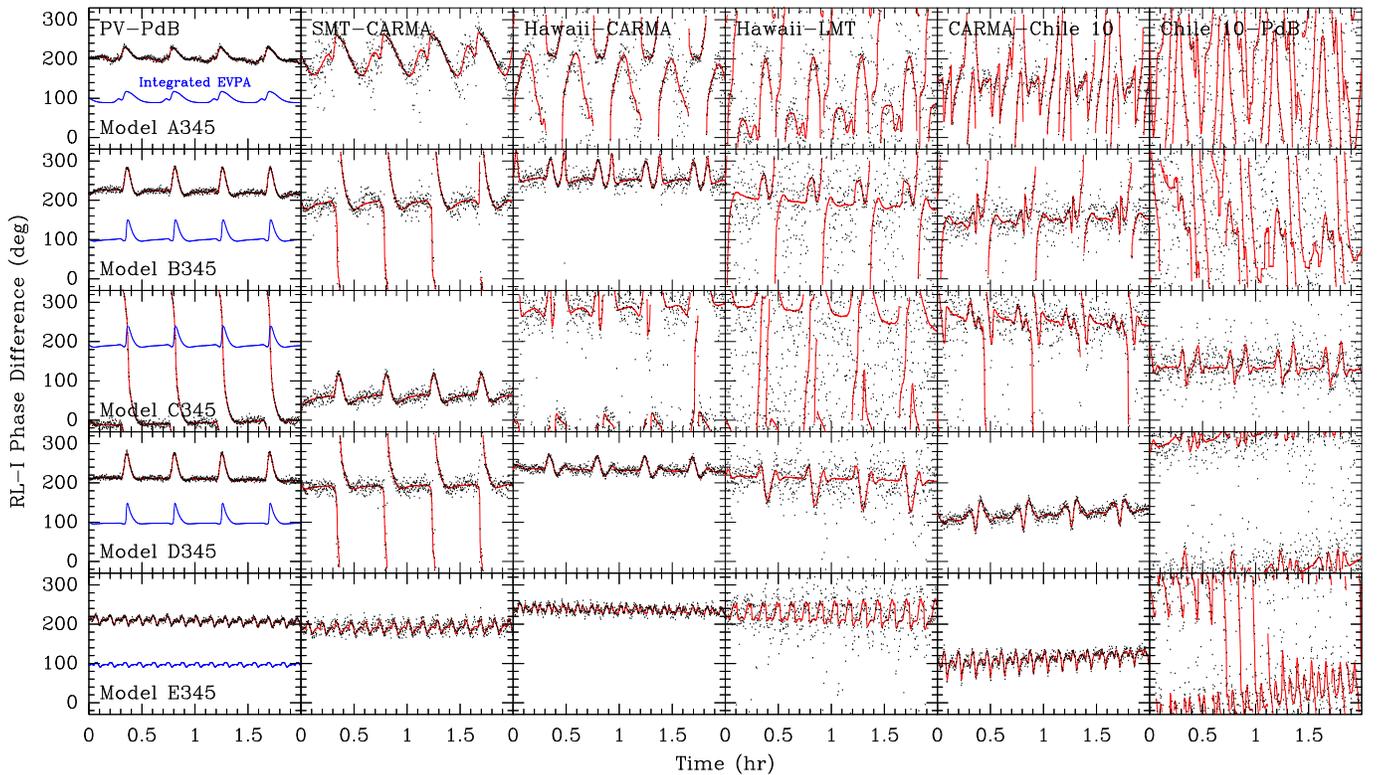}}
\caption{$RL-I$ phase differences for selected baselines at 345~GHz.
  See Figure~\ref{fig-phase230} for details.
\label{fig-phase345}
}
\end{figure*}

\begin{figure*}[p]
\resizebox{\hsize}{!}{\includegraphics{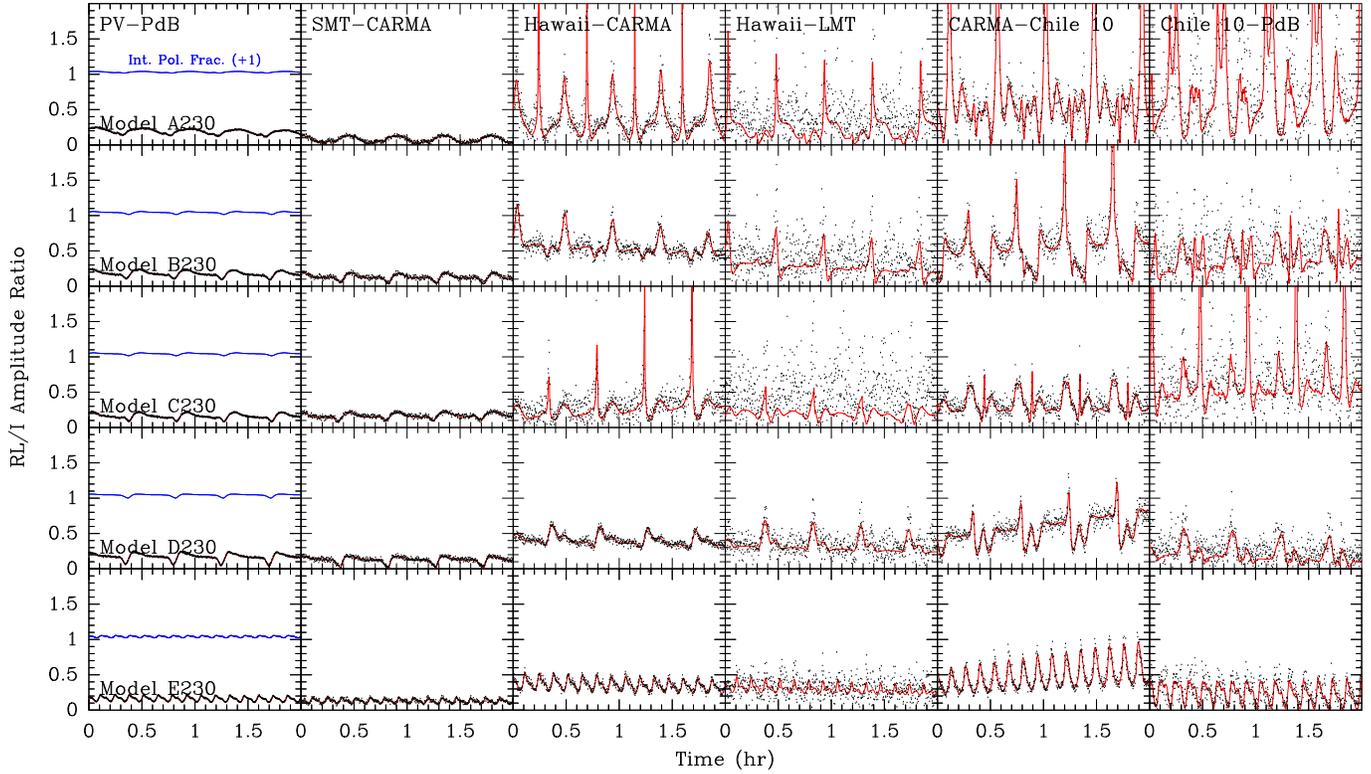}}
\caption{$RL/I$ amplitude ratio for selected baselines at 230~GHz.
  See Figure~\ref{fig-phase230} for details.  The blue line shows the
  integrated polarization fraction for the indicated model, shifted by
  1.0 for clarity.
\label{fig-amp230}
}
\end{figure*}

\begin{figure*}[p]
\resizebox{\hsize}{!}{\includegraphics{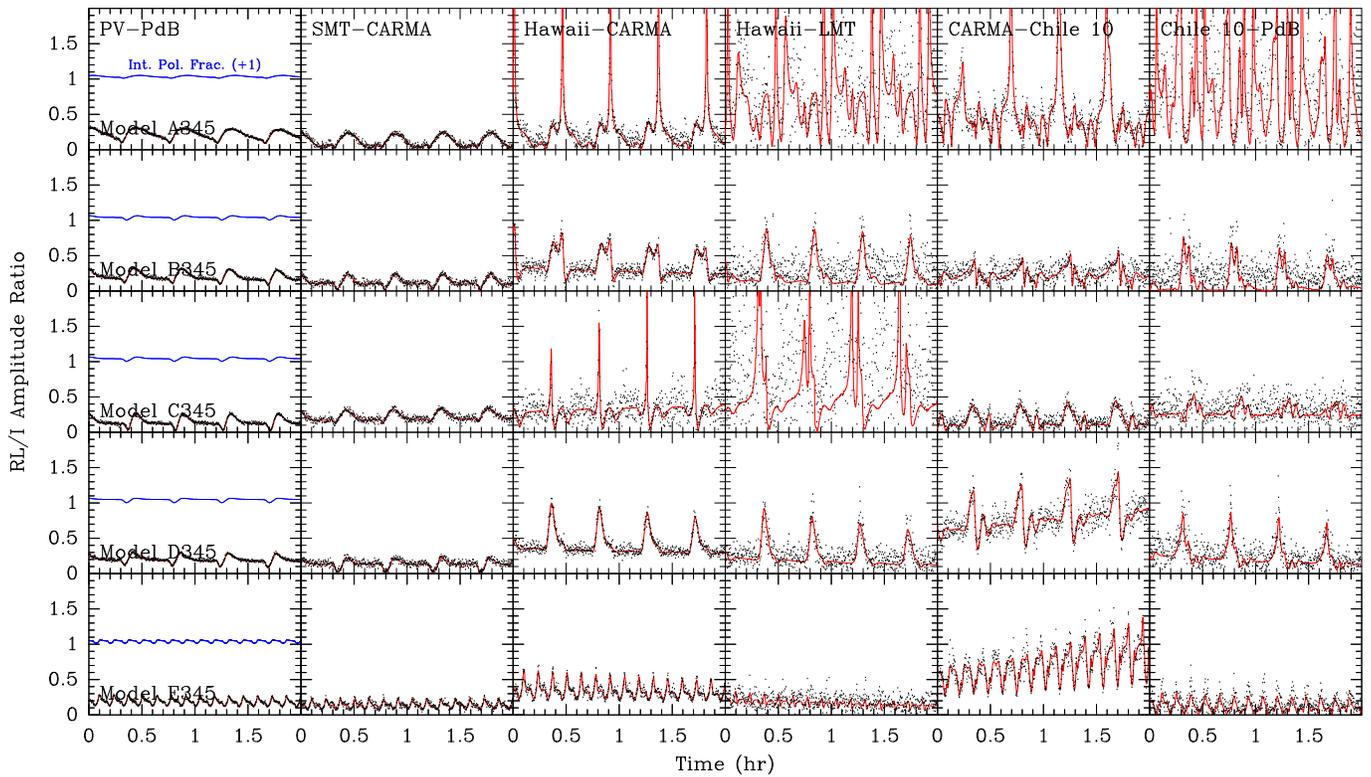}}
\caption{$RL/I$ amplitude ratio for selected baselines at 345~GHz.
  See Figures~\ref{fig-phase230} and \ref{fig-amp230} for details.
\label{fig-amp345}
}
\end{figure*}

\section{Results}
\subsection{Baseline Visibility Ratios}

\begin{figure}[t]
\resizebox{!}{6.0truein}{\includegraphics{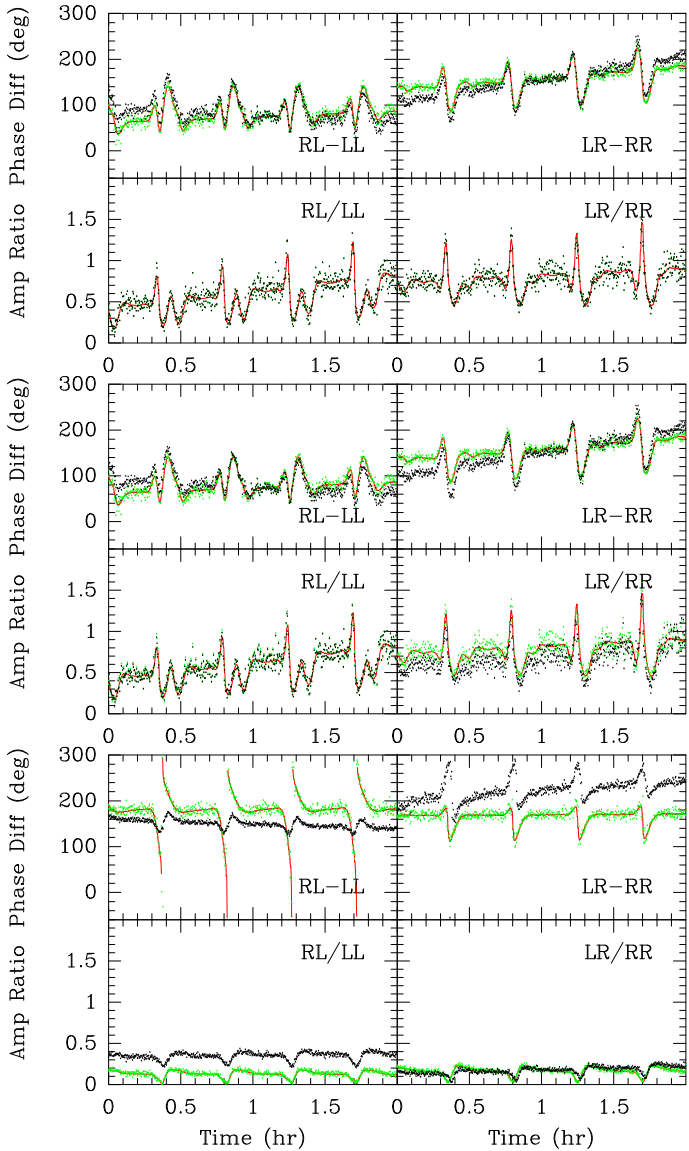}}
\caption{Example polarimetric visibility ratio quantities with and
  without parallactic angle and $D$-term calibration.  Noiseless model
  values are shown in red, simulated data (8~Gbit\,s$^{-1}$ per
  polarization) without parallactic angle and $D$-terms are shown in
  green, and simulated data with parallactic angle and $D$-terms are
  overplotted in black.  \emph{Top}: Simulated data for the
  CARMA-Chile~10 baseline of Model~D at 230~GHz with zero $D$-terms.
  Parallactic angle terms effectively produce a small slope in the
  phase difference terms over timescales of interest and have no
  effect on visibility amplitudes in the absence of $D$-terms (i.e.,
  the green and black dots are identical).  \emph{Middle}: The same
  baseline with $11 \pm 3$\% $D$-terms.  The inclusion of $D$-terms
  does not have a large effect because $|P|$ is several times larger
  than $|DI|$.  \emph{Bottom}: The SMT-CARMA baseline with $11 \pm
  3$\% $D$-terms.  Uncalibrated $D$-terms will bias amplitude ratios
  and may produce qualitatively different phase differences when the
  cross-to-parallel amplitude ratio is small but will not obscure
  periodicity even if the $D$-terms are large (as is assumed in this
  worst-case scenario based on the experience of \citet{attridge01}
  and \citet{attridge05} at 86~GHz).  The effect of uncalibrated
  $D$-terms will likely be largest on short baselines.
\label{fig-Dterms}
}
\end{figure}

While lower-resolution observations of Sgr~A* find polarization of
less than 10\%, the effective fractional polarization on smaller
scales can be much larger.  Figure~\ref{fig-uvplt} shows the
amplitudes of the $(u,v)$ data\footnote{It is important not to confuse
  the antenna spacing parameters measured in wavelengths
  (conventionally denoted by lower-case $u$ and $v$) with the Stokes
  parameters (denoted by upper-case $U$ and $V$).} that would be
produced by a disk and persistent, unchanging orbiting hot spot with
parameters as given in Model~A at 230~GHz.  The range in amplitudes
reflects the changing flux density, both in total flux (i.e., the
zero-spacing flux at $u = v = 0$) as well as on smaller spatial
scales, as would be sampled via VLBI.  Both total power (Stokes $I$)
and polarization signatures fall off with baseline length, but on
average the fractional polarization increases with longer baselines,
and the ratio of Stokes visibility amplitudes can exceed unity.  All
of our models produce much higher polarization fractions on small
angular scales than at large angular scales, and all models except for
Model~F at 345~GHz produce a substantial set of cross-to-parallel
visibility amplitude ratios in excess of unity on angular scales of
40--80~$\mu$as and smaller.

We henceforth focus on ratios of cross-to-parallel baseline
visibilities (e.g., $RL/I$).  Plots of the $RL-I$ phase
difference\footnote{Note that $\arg(RL) - \arg(I) = \arg(RL/I)$.} are
shown in Figures~\ref{fig-phase230} and \ref{fig-phase345} for models
at 230 and 345~GHz, respectively.  At a total data rate of
16~Gbit\,s$^{-1}$, nearly all baselines exhibit signatures of changing
polarization structure.  Due to the weak polarized signal on the
longest baselines, a phased array of a subset of ALMA (Chile~10, in
the nomenclature of \S \ref{models}) may be required in order to
confidently detect polarization changes on the long baselines,
especially to Europe.  The PV-PdB baseline (and to a lesser extent the
SMT-CARMA baseline at 230~GHz) effectively tracks the orientation of
the total linear polarization, since Sgr~A* is nearly unresolved on
this short baseline, and the calibrated $RL$ phase of a polarized
point source at phase center is twice the EVPA of the source
\citep[e.g.,][]{cotton95}.

Figures~\ref{fig-amp230} and \ref{fig-amp345} show the $RL/I$
visibility amplitude ratio for selected baselines at 230 and 345~GHz,
respectively.  The shortest baselines, PV-PdB and SMT-CARMA,
effectively track the large-scale polarization fraction as would be
measured by the SMA, for instance.  Because the short baselines
resolve out several tens of percent of the total intensity emission
(as compared to the zero-spacing flux in Fig.~\ref{fig-uvplt}) but a
much smaller fraction of the polarized emission, the variation in the
$RL/I$ and $LR/I$ amplitude ratios is fractionally larger than in the
large-scale polarization fraction.  A bias can be seen in the
amplitude ratios when the SNR is small, as noted in \S
\ref{polarimetric}.  (For brevity, we have shown only plots of the
$RL-I$ phase difference and $RL/I$ amplitude ratio.  The $LR-I$ phase
difference and $LR/I$ amplitude ratio exhibit similar behavior.)

Closure phases of the cross-hand terms can be constructed in the same
manner as for the parallel-hand terms, and these are robust
observables.  However, closure quantities are less necessary in the
polarimetric case than for total-intensity observations because robust
baseline-based observables can be constructed.  As
Figure~\ref{fig-uvplt} shows, the visibility amplitude in the
cross-hand correlations is much lower than that of the parallel-hand
correlations on short baselines.  The SNR of the closure phase is
lower by a factor of $\sqrt{3}$ than the three constituent baseline
SNRs when the latter are all equal and is dominated by that of the
weakest baseline when there is a large difference in the baseline SNRs
\citep{rogers95}.  In Stokes $I$, the mean baseline SNRs (averaged
over multiple orbits) are greater than or equal to 5 on virtually all
baselines and all models at 16~Gbit\,s$^{-1}$ total bit rate
(8~Gbit\,s$^{-1}$ each RR and LL) in a 10~s coherence interval,
provided that the Chile~10 is used in lieu of Chile~1.  The number of
triangles with SNRs greater than 5 on all baselines at
8~Gbit\,s$^{-1}$ in $RL$ or $LR$ is much smaller.  Depending on the
model, the SMT-CARMA-LMT and SMT/CARMA-LMT-Chile~10 triangles usually
satisfy this condition, with Hawaii-SMT-CARMA also having sufficient
SNR.  Completion of the LMT, resulting in a system equivalent flux
density of $\lesssim 600$~Jy at 230~GHz, will allow for strong
detections on the Hawaii-LMT baseline and, importantly, significantly
strengthen detections on the LMT-Chile baseline.  If the coherence
time is significantly shorter than 10~s, or if the obseved flare flux
density is substantially lower than assumed in our models, closure
phases may not have a large enough SNR to detect periodic changes.  In
any case, if polarimetric visibility ratios are successful in
detecting periodicity, there may not be a need to appeal to closure
quantities except insofar as they can be used to improve the array
calibration.

\subsection{Instrumental Polarization Calibration}

We have also simulated the effects of not correcting for parallactic
angle terms and instrumental polarization by including Gaussian random
$D$-terms of $11 \pm 3$\% with uniform random phases, based on the
\citet{attridge01} and \citet{attridge05} CMVA studies.  This should
be considered a worst-case scenario.  $D$-terms at many of the
telescopes will likely be substantially better: e.g., 1-6\% at the SMA
in observations by \citet{marrone06a,marrone07} and about 5\% at the 6
m antennas of the CARMA array \citep{bower02}.  These quantities
affect the observed correlation quantities $RR$, $LL$, $RL$, and $LR$
as indicated in \S \ref{polarimetric}.  We have ignored terms of order
$D^2$, but we have included terms of the form $D \, P$, since the
polarized visibility amplitudes can be larger than the Stokes $I =
\onehalf \,(RR+LL)$ visibility amplitudes on long baselines
(Fig.~\ref{fig-uvplt}).

Example data showing the effects of large uncalibrated $D$-terms is
shown in Figure~\ref{fig-Dterms}.  Instrumental polarization adds a
bias to the ratio of cross-hand to parallel-hand visibility amplitudes
(e.g., $RL/RR$) as well as a phase slope and offset to the difference
of cross-hand and parallel-hand phases (e.g., $RL-RR$).  These effects
are much more pronounced on the short baselines, especially PV-PdB and
SMT-CARMA, because the fractional source polarization on large scales
is small (and thus $|D\,I| \not \ll |P|$).  In most cases, the
cross-to-parallel amplitude ratios and phase differences behave
similarly whether instrumental polarization calibration is included or
not simply by virtue of the fact that the polarized intensity is a
large fraction of the total intensity.  Deviations in the
cross-to-parallel phase difference response appear qualitatively large
when the cross-hand amplitudes are near zero because small offsets
from the source visibility, represented as a vector in the complex
plane, can produce large changes in the angle (i.e., phase) of the
visibility.  Large instrumental polarization can affect the expected
baseline-based signatures but do not obscure periodicity, since source
structure changes in Stokes $I$ and $P$ have the same period in our
models.  Of course, proper $D$-term calibration is a sine qua non for
modelling the polarized source structure (but not for detecting
periodicity).  The $D$-terms can be measured by observing a bright
unpolarized calibrator (or polarized, unresolved calibrator), and the
visibilities should be corrected for instrumental polarization if
possible.

\subsection{Periodicity Detection}

\begin{figure*}
\resizebox{\hsize}{!}{\includegraphics{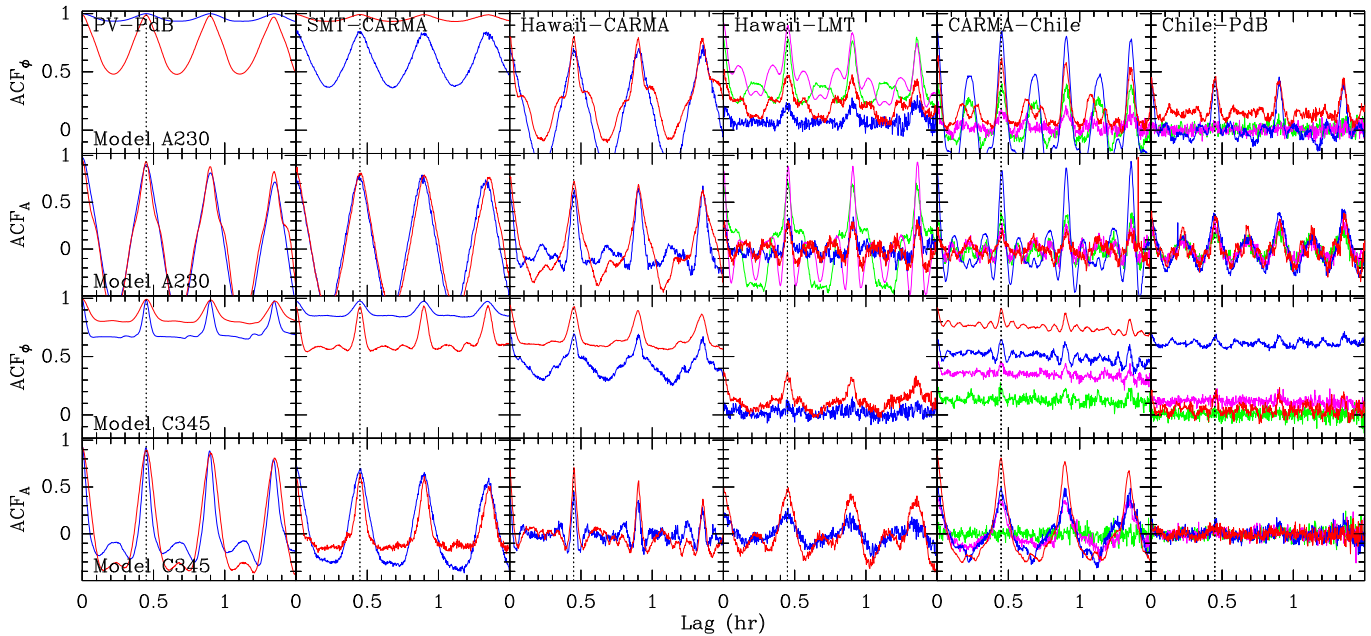}}
\caption{Autocorrelation functions for models A230 and C345 on
  selected baselines.  Blue and red denote $RL/I$ and $LR/I$,
  respectively.  Green and magenta indicate $RL/I$ and $LR/I$,
  respectively, using a completed 50~m LMT optimized for 230~GHz
  performance (fourth column) or Chile~1 (in place of Chile~10 in the
  fifth and sixth columns).  The dashed line indicates the orbital
  period.
\label{fig-acfs}
}
\end{figure*}

As in Paper I, we can define autocorrelation functions to test for
periodicity.  More optimal methods exist to extract the period of a
time series of data \citep{rogers09}, but the autocorrelation function
is conceptually simple and suffices for our models.  The amplitude
autocorrelation function evaluated at lag $k$ on a time series of $n$
amplitude ratios $A_i = RL_i/I_i$ (or $LR_i/I_i$) on a baseline is
defined as
\[
\mathrm{ACF}_A(k) \equiv \frac{1}{(n-k)\sigma^2}
\sum\limits_{i=1}^{n-k}\left[(\log A_i - \mu)(\log A_{i+k} - \mu)\right],
\]
where $\mu$ and $\sigma^2$ are the mean and variance of the logarithm
of the amplitude ratios, respectively.  The phase autocorrelation
function is defined as
\[
\mathrm{ACF}_\phi(k) \equiv \frac{1}{n-k}
\sum\limits_{i=1}^{n-k}\cos(\phi_i - \phi_{i+k}),
\]
where $\phi_i$ denotes the $RL-I$ or $LR-I$ phase difference of point
$i$.  By definition, $\mathrm{ACF}_A(0) = \mathrm{ACF}_\phi(0) = 1$.
The largest non-trivial peak corresponds to the period, with the
caveat that the changing baseline geometries caused by Earth rotation
can conspire to cause the autocorrelation function to be slightly
greater at integer multiples of the true period.  The phase
autocorrelation function can suffer from lack of contrast when the
visibility phase difference is not highly variable as may be the case
for the shortest baselines depending on the model
(Fig.~\ref{fig-acfs}), but the lack of contrast is not so severe as in
the total-intensity case (cf.\ Paper I) due to the sensitivity of
short-baseline cross-to-parallel phase differences to the
source-integrated EVPA.

An array consisting of Hawaii, SMT, and CARMA is sufficient to
confidently detect periodicity at a total bit rate of 2~Gbit\,s$^{-1}$
(i.e., 0.25~GHz bandwidth per polarization) over 4.5 orbits of data
for Models~A-E.  This contrasts with the total intensity case, in
which a substantially higher bit rate is required on the same array,
depending on the model (Paper I).  The key point is that the
cross-to-parallel visibility ratios on the short baselines trace the
overall source polarization fraction and EVPA, which are readily
apparent even at the much coarser angular resolution afforded by
connected-element interferometry \citep[e.g.,][]{marrone06b}.  Long
baselines are thus not strictly necessary to detect periodic
polarimetric structural changes, although they will be important for
modelling the small-scale polarimetric structure of the Sgr~A*.  In
contrast, significantly higher bit rates and 4-element arrays are
usually required to detect periodic source structure changes in total
intensity (Paper I).

Long-period models (e.g., Model~F) are problematic for millimeter VLBI
periodicity detection because it may not be possible to detect more
than two full periods during the window of mutual visibility between
most of the telescopes in a potential VLBI array.  As with the total
intensity case (Paper~I), the most promising approach for millimeter
VLBI is to observe with an array of four or five telescopes, since
large changes in cross-to-parallel phase differences and visibility
amplitude ratios tend to be episodic across most or all baselines.
The LMT is usefully placed because it provides a long window of mutual
visibility to Chile and the US telescopes as well as a small time
overlap with PV, thus enabling a large continuous time range over
which Sgr~A* is observed.  Assuming that the flare source can survive
for several orbital periods, connected-element interferometry may
suffice to demonstrate periodicity.  Sgr~A* is above
10\degr\ elevation for approximately 9~hr from Hawaii and 12~hr from
Chile.

\section{Discussion}

\subsection{Faraday Rotation and Depolarization}
\label{faraday}

VLBI polarimetry has several key advantages over single-dish and
connected-element interferometry for understanding the polarization
properties of Sgr~A*.  First, even the shortest baselines likely to be
included in the array will filter out surrounding emission.  Reliable
single-dish extraction of polarization information requires
subtracting the contribution from the surrounding dust, which can
dominate the total polarized flux at 345~GHz and is significant even
at 230~GHz \citep{aitken00}.  Contamination by surrounding emission is
much less severe for SMA measurements, where the synthesized beamsize
is on the order of an arcsecond, depending on configuration \citep[for
  instance, $1\farcs4 \times 2\farcs2$ in the observations
  of][]{marrone06a}.  VLBI will do much better still, with the
shortest baselines resolving out most of the emission on scales larger
than $\sim 1$~mas ($100~R_S$), effectively restricting sensitivity to
the inner accretion disk and/or outflow region.

Second, the resolution provided by millimeter VLBI will greatly reduce
depolarization due to blending of emission from regions with different
linear polarization directions.  Models of the accretion flow predict
that linear polarization position angles and Faraday rotation will be
nonuniform throughout the source
\citep{bromley01,broderick05,broderick06,huang08}.  For this reason,
ratios of cross-hand to parallel-hand visibilities (which are the
visibility analogues of linear polarization fractions) can greatly
exceed the total linear polarization fraction integrated over the
source (cf.\ Figure~\ref{fig-uvplt} and Table~\ref{tab-models}).

Third, VLBI polarimetry has the potential to identify whether changes
in detected polarization are due to intrinsic source variability or
changes in the rotation measure at larger distances.  The former would
be expected to be variable on relatively short timescales (minutes to
tens of minutes), consistent with the orbital period of emission at a
few gravitational radii.  The latter would be expected to vary more
slowly and affect only the polarized emission, not the total
intensity.  A cross-correlation between polarized and total-intensity
data may allow the two effects to be disentangled.

Linear polarization at millimeter wavelengths can be used to estimate
the accretion rate of Sgr~A* \citep[e.g.,][]{quataert00}.  At
frequencies below $\sim 100$~GHz, no linear polarization is detected
due to Faraday depolarization in the accretion region
\citep{bower99a,bower99b}.  Linear polarization is detected toward
Sgr~A* at higher frequencies \citep[e.g.,]{aitken00}, where the
effects of Faraday rotation are smaller.  Ultimately, accretion rates
are constrained by the lack of linear polarization at long wavelengths
and its existence at short wavelengths.  Measurements of the Faraday
rotation exist, although it is unclear whether changes in detected
polarization angles are due to changing source polarization structure
or a variable rotation measure \citep{bower05,marrone06a,marrone07}.

Longer term, imaging may be possible if all seven millimeter telescope
sites heretofore considered (and possibly others as well) are used
together as a global VLBI array \citep[e.g., an Event Horizon
  Telescope; ][]{doelemaneht}.  Imaging the quiescent polarization
structure of Sgr~A* may allow the characteristics of the source
emission region to be distinguished from those of the region producing
Faraday rotation (which may overlap or be identical with the emission
region).  Contemporaneous millimeter VLBI observations at two
different frequencies would allow separate maps of the intrinsic
polarization structure and the rotation measure to be produced.  It
may also then be possible to place strong constraints on the density
($n_e$) and magnetic fields ($B$) in Sgr~A*.  Briefly, the rotation
measure is related to $\int n_e \, B_\parallel dl$, while the total
intensity is proportional to $\int n_e \, B_\perp^{\alpha+1} dl$,
where $\alpha$ is the optically thin spectral index
\citep{westfold59}.  Obtaining these results will require the ability
to fully calibrate the data for instrumental polarization terms and
the absolute EVPA.  It may also require higher image fidelity than a
seven-telescope VLBI array can provide \citep{fish09}.  There are
possibilities for extending a millimeter VLBI array beyond these seven
sites by adding other existing (e.g., the South Pole Telescope) or new
telescopes \citep{doelemaneht}, but full consideration of the
scientific impact of potential future arrays is beyond the scope of
this work.

\subsection{Physical Considerations}
\label{physical}

The inclusion of a screen of constant Faraday rotation alters the
phases of the cross-hand terms (and therefore the cross-to-parallel
phase differences) but does not materially affect the detectability of
changing polarization structure.  The mean rotation measure of Sgr~A*
averaged over multiple epochs is $(-5.6 \pm 0.7) \times
10^5$~rad\,m$^{-2}$ \citep{marrone07}, which corresponds to a rotation
of polarization vectors by $-55\degr$ at 230~GHz and $-24\degr$ at
345~GHz.  Persistent gradients of rotation measure across the source
are virtually indistinguishable from intrinsic polarization structure
in the case of a steady-state source, but it is possible that the
source structure and rotation measure change on different timescales,
which would allow the two effects to be disentangled
\citep{marrone07}.  Comparison of changes in the polarization data
with total intensity data (obtained from the parallel-hand
correlations) and total polarization fraction (obtained from
simultaneous connected-element interferometric data if available, else
inferred from the shortest VLBI baselines) may be useful for
identifying whether observed polarization angle changes are due to a
variable rotation measure \citep{bower05}.

The physical mechanism that produces flares in total intensity and
polarization changes is poorly understood.  Connected-element
interferometry at millimeter wavelengths has not been conclusive as to
whether orbiting hot spots are the underlying mechanism that produces
flares in Sgr~A* (cf.\ \citealt{marrone06b} and \citealt{marrone08}),
or even as to whether multiple mechanisms may be responsible for
flaring.  Polarization variability can be decorrelated from total
intensity variability \citep[e.g.,][]{marrone06b}, and each shows
variability on time scales ranging from tens of minutes to hours (and
possibly longer).  Spatial resolution will be key to deciphering the
environment of Sgr~A*, and thus there is a critical need for
polarimetric millimeter-wavelength VLBI.

Our results are generalizable to any mechanism producing changes in
linear polarization, whether due to orbiting or spiralling hot spots,
jets, disk instabilities, or any other mechanism in the inner disk of
Sgr~A*.  Visibility ratios on baselines available for
millimeter-wavelength VLBI will provide reasonably robust observables
to detect changes in the polarization structure on relevant scales
from a few to a few hundred $R_\mathrm{S}$, regardless of the cause of
those changes.  Clearly, periodicity can only be detected if the
underlying mechanism that produces polarization changes is itself
periodic, but baseline visibility ratios will be sensitive to any
changes that are rapid compared to the rotation of the Earth.

\subsection{Observational Strategy}

Future millimeter-wavelength VLBI observations of Sgr~A* should
clearly be observed in dual-polarization unless not allowed by
telescope limitations.  Total-intensity analysis via closure
quantities, as outlined in Paper I, can be performed regardless of
whether the data are taken in single- or dual-polarization mode, but
the cross-hand correlations can only be obtained from
dual-polarization data.  The cross-hand correlation data provide
additional chances to detect variability via changing source
polarization structure.

By virtue of its size and location, which produces medium-length
baselines to Chile and Hawaii as well as a long window of mutual
visibility with Chile, the LMT is a very useful telescope.  The
sensitivities assumed in this work for first light on the LMT may lead
to biased amplitude ratios on the Hawaii-LMT baseline
(Fig.~\ref{fig-amp230}), but this will not prevent detection of
periodicity (Fig.~\ref{fig-acfs}).  Thus, strong consideration should
be given in favor of including the LMT in a millimeter-wavelength VLBI
array observing Sgr~A* as soon as possible.  If the parameters of the
fully-completed LMT are assumed, the bias disappears and the scatter
of points on the Hawaii-LMT baseline in Figure~\ref{fig-amp230} is
similar to that seen on the shorter Hawaii-CARMA baseline, and
baselines between continental North America and the LMT will be of
comparably good SNR to the lower-resolution PV-PdB baseline.
Eventually, the LMT and ALMA will be the most sensitive stations in a
millimeter VLBI array and will enable sensitive modelling of the
Sgr~A* system.

If possible, it would be advantageous to obtain connected-element
interferometric data of Sgr~A* simultaneously with VLBI data.  While
amplitude ratios and phase differences on the PV-PdB and (to a lesser
extent) SMT-CARMA baselines track the large-scale polarization
fraction and EVPA fairly well in these models, it is not known what
fraction of the polarization structure arises from larger-scale
emission in Sgr~A*.  If interferometer stations can be configured to
produce both cross-correlations betweeen telescopes as well as a
phased output of all telescopes together, opportunities for
simultaneous connected-element interferometry may exist with the PdB
Interferometer, the SMA, CARMA, or ALMA.  If system limitations
prevent this, it may still be possible to acquire very-short-spacing
data with those telescopes in CARMA or ALMA that are not phased
together for VLBI.

\section{Conclusions}

(Sub)millimeter-wavelength VLBI polarimetry is a very valuable
diagnostic of emission processes and dynamics near the event horizon
of Sgr~A*.  We summarize the findings in this paper as follows:

\begin{itemize}
\addtolength{\itemsep}{-0.8ex}
\item Millimeter-wavelength polarimetric VLBI can detect changing
  source structures.  Despite low polarization fractions seen with
  connected-element interferometry, the much higher angular resolution
  data provided by VLBI will be far less affected by beam
  depolarization and contamination from dust polarization.
  Polarimetric VLBI provides an orthogonal way to detect periodic
  structural changes as compared with total intensity VLBI.

\item Ratios of cross- to parallel-hand visibilities are robust
  baseline-based observables.  Short VLBI baselines approximately
  trace the integrated polarization fraction and position angle of the
  inner accretion flow of Sgr~A*, while longer VLBI baselines resolve
  smaller structures.

\item Calibration of instrumental polarization terms is not necessary
  to detect a changing source structure, including periodicity, in
  Sgr~A*.

\item Polarimetric VLBI may be able to disentangle the effects of
  rotation measure from intrinsic source polarization.  Initial
  results will likely come from observations of the timescale of
  polarimetric variability.  If the initial array is expanded to allow
  high-fidelity imaging, polarimetric VLBI may be able to map the
  Faraday rotation region and directly infer the density and magnetic
  field structure of the emitting region in Sgr~A*.

\end{itemize}

\acknowledgments

The high-frequency VLBI program at Haystack Observatory is funded
through a grant from the National Science Foundation.


\begin{thebibliography}{}

\bibitem[Aitken et al.(2000)]{aitken00} Aitken, D.~K., Greaves, J.,
  Chrysostomou, A., Jenness, T., Holland, W., Hough, J.~H.,
  Pierce-Price, D., \& Richer, J.\ 2000, \apjl, 534, L173

\bibitem[Attridge(2001)]{attridge01} Attridge, J.~M.\ 2001, \apjl,
  553, L31

\bibitem[Attridge et al.(2005)]{attridge05} Attridge, J.~M., Wardle,
  J.~F.~C., \& Homan, D.~C.\ 2005, \apjl, 633, L85

\bibitem[Bower et al.(1999a)]{bower99a} Bower, G.~C., Backer, D.~C.,
  Zhao, J.-H., Goss, M., \& Falcke, H.\ 1999a, \apj, 521, 582

\bibitem[Bower et al.(2005)]{bower05} Bower, G.~C., Falcke, H.,
  Wright, M.~C.~H., \& Backer, D.~C.\ 2005, \apjl, 618, L29

\bibitem[Bower et al.(1999b)]{bower99b} Bower, G.~C., Wright, M.~C.H.,
  Backer, D.~C., \& Falcke, H.\ 1999b, \apj, 527, 851

\bibitem[Bower et al.(2003)]{bower03} Bower, G.~C., Wright, M.~C.~H.,
  Falcke, H., \& Backer, D.~C.\ 2003, \apj, 588, 331

\bibitem[Bower et al.(2002)]{bower02} Bower, G.~C., Wright, M.~C.~H.,
  \& Forster, R.\ 2002, Polarization Stability of the BIMA Array at
  1.3~mm, BIMA memo 89

\bibitem[Broderick \& Blandford(2004)]{broderick04} Broderick, A., \&
Blandford. R.\ 2004, \mnras, 349, 994

\bibitem[Broderick \& Loeb(2005)]{broderick05} Broderick, A.~E., \&
  Loeb, A.\ 2005, \mnras, 363, 353

\bibitem[Broderick \& Loeb(2006)]{broderick06} Broderick, A.~E., \&
  Loeb, A.\ 2006, \mnras, 367, 905

\bibitem[Bromley et al.(2001)]{bromley01} Bromley, B.~C., Melia, F.,
  \& Liu, S.\ 2001, \apjl, 555, L83

\bibitem[Brown et al.(1989)]{brown89} Brown, L.~F., Roberts, D.~H., \&
  Wardle, J.~F.~C.\ 1989, \aj, 97, 1522

\bibitem[Cotton(1993)]{cotton93} Cotton, W.~D.\ 1993, \aj, 106, 1241

\bibitem[Cotton(1995)]{cotton95} Cotton, W.~D.\ 1995, Very Long
  Baseline Interferometry and the VLBA, 82, 289

\bibitem[Doeleman et al.(2009a)]{doeleman09} Doeleman, S.~S., Fish,
  V.~L., Broderick, A.~E., Loeb, A., \& Rogers, A.~E.~E.\ 2009a, \apj,
  695, 59

\bibitem[Doeleman et al.(2008)]{doeleman08} Doeleman, S.~S., et
  al.\ 2008, \nat, 455, 78

\bibitem[Doeleman et al.(2009b)]{doelemaneht} Doeleman, S.~S., et
  al.\ 2009b, Astronomy, 2010, 68

\bibitem[Fish \& Doeleman(2009)]{fish09} Fish, V.~L., \& Doeleman,
  S.~S.\ 2009, IAU Symposium, 261, submitted, arXiv:0906.4040

\bibitem[Huang et al.(2008)]{huang08} Huang, L., Liu, S., Shen, Z.-Q.,
  Cai, M.~J., Li, H., \& Fryer, C.~L.\ 2008, \apj, 676, L119

\bibitem[Huang et al.(2009)]{huang09} Huang, L., Liu, S., Shen, Z.-Q.,
  Yuan, Y.-F., Cai, M.~J., Li, H., \& Fryer, C.~L.\ 2009, \apj, 703,
  557

\bibitem[Jones \& O'Dell(2007)]{jones77} Jones, T.~W., \& O'Dell,
  S.~L.\ 1977, \apj, 214, 522

\bibitem[Macquart et al.(2006)]{macquart06} Macquart, J.-P., Bower,
  G.~C., Wright, M.~C.~H., Backer, D.~C., \& Falcke, H.\ 2006, \apjl,
  646, L111

\bibitem[Markoff et al.(2007)]{markoff07} Markoff, S., Bower, G.~C.,
  \& Falcke, H.\ 2007, \mnras, 379, 1519

\bibitem[Marrone et al.(2006a)]{marrone06a} Marrone, D.~P., Moran,
  J.~M., Zhao, J.-H., \& Rao, R.\ 2006a, \apj, 640, 308

\bibitem[Marrone et al.(2006b)]{marrone06b} Marrone, D.~P., Moran, 
  J.~M., Zhao, J.-H., \& Rao, R.\ 2006b, Journal of Physics Conference
  Series, 54, 354

\bibitem[Marrone et al.(2007)]{marrone07} Marrone, D.~P., Moran,
  J.~M., Zhao, J.-H., \& Rao, R.\ 2007, \apjl, 654, L57

\bibitem[Marrone et al.(2008)]{marrone08} Marrone, D.~P., et al.\
  2008, \apj, 682, 373

\bibitem[Petrosian \& McTiernan(1983)]{petrosian83} Petrosian, V., \&
  McTiernan, J.~M.\ 1983, Phys. Fluids, 26, 3023

\bibitem[Quataert \& Gruzinov(2000)]{quataert00} Quataert, E., \&
  Gruzinov, A.\ 2000, \apj, 545, 842

\bibitem[Roberts et al.(1994)]{roberts94} Roberts, D.~H., Wardle,
  J.~F.~C., \& Brown, L.~F.\ 1994, \apj, 427, 718

\bibitem[Rogers et al.(2009)]{rogers09} Rogers, A.~E.~E., Doeleman,
  S.~S., \& Fish, V.~L.\ 2009, \baas, 41, 217

\bibitem[Rogers et al.(1995)]{rogers95} Rogers, A.~E.~E., Doeleman,
  S.~S., \& Moran, J.~M.\ 1995, \aj, 109, 1391

\bibitem[Rogers et al.(1974)]{rogers74} Rogers, A.~E.~E., et
  al.\ 1974, \apj, 193, 293

\bibitem[Thompson et al.(2001)]{thompson01} Thompson, A.~R., Moran,
  J.~M., \& Swenson, G.~W, Jr.\ 2001, Interferometry and Synthesis in
  Radio Astronomy (2nd ed; New York: Wiley)

\bibitem[Trippe et al.(2007)]{trippe07} Trippe, S., Paumard, T., Ott,
  T., Gillessen, S., Eisenhauer, F., Martins, F., \& Genzel, R.\ 2007,
  \mnras, 375, 764

\bibitem[Wardle(1971)]{wardle71} Wardle, J.~F.~C.\ 1971, \aplett, 8,
  53

\bibitem[Westfold(1959)]{westfold59} Westfold, K.~C.\ 1959, \apj, 130,
  241

\bibitem[Yuan et al.(2003)]{yuan03} Yuan, F., Quataert, E., \&
  Narayan, R.\ 2003, \apj, 598, 301

\end{thebibliography}
\end{document}